\documentstyle[eqsecnum,multicol,aps,prl,epsf]{revtex}

\def\ben{\begin{equation}}
\def\een{\end{equation}}
\def\bea{\begin{eqnarray}}
\def\eea{\end{eqnarray}}

\date{\today}
\draft
\tighten
\begin{document}
\def\sqr#1#2{{\vcenter{\hrule height.3pt
      \hbox{\vrule width.3pt height#2pt  \kern#1pt
         \vrule width.3pt}  \hrule height.3pt}}}
\def\square{\mathchoice{\sqr67\,}{\sqr67\,}\sqr{3}{3.5}\sqr{3}{3.5}}
\def\today{\ifcase\month\or
  January\or February\or March\or April\or May\or June\or July\or
  August\or September\or October\or November\or December\fi
  \space\number\day, \number\year}

\def\Bbb{\bf}


\title{Supergravity on the Brane}

\author{A. Chamblin$^{*}$\thanks{$^{*}$Address from 15 October, 1999:
Center for Theoretical Physics, MIT, Bldg. 6-304,
77 Massachusetts Ave., Cambridge, MA 02139, USA} \& G.W. Gibbons}

\address {\qquad \\ DAMTP,
Silver Street,
Cambridge, CB3 9EW, England
}
\maketitle


\begin{abstract} 
We show that smooth domain wall spacetimes supported by a scalar field 
separating two anti-de-Sitter like regions admit a single graviton
bound state. Our analysis yields a fully 
non-linear supergravity treatment of the Randall-Sundrum model.
Our solutions describe
a pp-wave propagating in the domain wall background spacetime.  
If the latter is BPS, our solutions retain some supersymmetry.
Nevertheless, the Kaluza-Klein modes generate
``pp curvature'' singularities in the bulk
located where the horizon of AdS would ordinarily be.   
\end{abstract}

\pacs{12.10.-g, 11.10.Kk, 11.25.M, 04.50.+h 
{\qquad} {\qquad} {\qquad} {\qquad} {\qquad} {\qquad}
{\qquad} {\qquad}  DAMTP-1999-126}
\vspace*{0.1cm}

\begin{multicols} {2}

\section{Introduction}
It has long been thought that any attempt to model the Universe as a 
single brane
embedded in a higher-dimensional bulk spacetime must inevitably
fail because
the gravitational forces experienced by matter on the brane,
being mediated by gravitons travelling in the bulk, are those
appropriate to the higher dimensional spacetime rather than the
lower dimensional brane. Recently however, Randall and Sundrum
have argued that there are circumstances under which this need not be so.
Their model involves a thin ``distributional" static flat
domain wall or three-brane
separating two regions of five-dimensional anti-de-Sitter spacetime.
They solve for the linearized graviton perturbations and find a 
square integrable bound state representing a gravitational
wave confined to the domain wall. They also found the linearized
bulk or ``Kaluza-Klein" graviton modes. They argue that
the latter decouple from the brane and 
make negligible contribution to the force beween two
sources in the brane, so that this force is due primarily to the 
bound state.  In this way we get an inverse square law
attraction rather than the inverse cube law one might 
naively have anticipated (see \cite{gog} for a related discussion).

This result is rather striking and raises various questions.
For example one would like to 
know how general the effect is.  Is it just an effect of the 
linearized perturbations or does it persist when 
non-linearities are taken into account?
One would
expect to get only one massless spin two bound state 
if the effective theory on the brane is to be general relativity.
In their derivation a crucial role
is played by a delta-function in the linearized graviton equation of 
motion. This is responsible for the unique bound state. 
It also seems that the effect will only work for domain walls and not for 
other branes. However the full dynamics
of the domain wall  is not treated in detail in the Randall-Sundrum  model.
In fact gravitating domain walls have a drastic effect on 
the curvature of the ambient spacetime and it is not obvious that a simple model involving a 
single collective coordinate representing the transverse 
displacement of the domain wall is valid. 

For these reasons it seems desirable to have a simple non-singular
model which is exactly solvable.
It is the purpose of this note to provide that.

\section{Thick domain walls in adS}

We first seek a static domain wall solution of of the d-dimensional
Einsten equations 
\begin{eqnarray}
R_{mn} -{ 1\over 2} R g_{mn} =  
\partial _m \Phi \cdot  \partial _n \Phi \\
- g_{mn} 
\Bigl ( { 1\over 2} \partial _a \Phi \cdot \partial _b \Phi\ g^{ab} + 
V(\Phi) \Bigr ) \nonumber
\end{eqnarray}

\noindent where  $a,b=0,1,2,\dots, d-1$.
The right hand side of (2.1)
is the energy momentum tensor of one or more scalar fields $\Phi$
with potential $V(\Phi)$ whose
kinetic energy term may contain a non-trivial metric on the 
scalar field manifold.  The metric is assumed to be of the form:
\ben
ds^2= dr^2 + e^{2A(r)} \eta_{\mu \nu} dx^\mu dx^\nu \label{metI},
\een
where $\mu, \nu =0,1,2,\dots, d-2$ and $\eta_{\mu \nu}$ is the flat 
Minkowski metric.
The scalar field is assumed to depend only on the transverse
coordinate $r$ and if $\prime$ denotes differentation with respect to $r$
then the Einstein equations require 
\begin{eqnarray}
-\Phi^\prime \cdot \Phi^\prime &=& (d-2)A^{\prime \prime},\\ 
\Bigl ({ 1\over 2} \Phi^\prime \cdot \Phi^\prime-V \Bigr ) &=&
{(d-2)(d-1) \over 2} (A^{\prime}) ^2.\nonumber
\end{eqnarray}
These two equations imply the scalar field equation:
\ben
\Phi ^{\prime \prime } + (d-1) \Phi ^\prime A^\prime = {\partial V \over \partial \Phi} \label{Scal}.
\een
If 
there is a non-trivial covariant metric on the scalar field 
manifold the right hand side of (\ref{Scal}) includes the contravariant
metric.

A domain wall solution separating two 
anti-de-Sitter domains with the same cosmological constant would have
has $A \approx -|r|/a $ as $|r| \rightarrow \infty$.

If the potential $V$ has the special form
\ben
V= {1 \over 2} \Bigl ( {{\partial} W \over 
{\partial} \Phi } \cdot {{\partial} W \over 
{\partial} \Phi } - { d-1 \over d-2} W^2 \Bigr ) 
\een
where $W = W(\Phi)$ is a suitable superpotential 
then Einstein equations (2.3) and the scalar equation 
(\ref{Scal}) are solved by solutions of the first order Bogomol'nyi
equations:
\ben
\Phi ^\prime =  {{\partial} W  \over {\partial} \Phi},
\thinspace \thinspace \thinspace
A^\prime =-{1 \over d-2} W, \label{flow}
\een

Note that the spacetime is uniquely specified by giving a solution
of (\ref{flow}) which is the same as the equation for a 
domain wall in the absence of gravity. One then obtains $A$ by 
quadratures. The vacua correspond to critical points of the
superpotential $W$. At these points the potential $V$ is negative,
and so one is in an anti-de-Sitter phase.  Recently, there 
has been a lot of interest in the possibility of obtaining such
potentials within the context of $d = 5$ gauged 
supergravity models (\cite{cvetic},\cite{freedman}, 
\cite{skenderis}, \cite{kallosh}. At present no
superpotential with the correct properties derived
from a supergravity model has yet been found.  
However a solution was exhibited in \cite{wfgk}
which is not derived from a supergravity model.
We will return to this point in the last section. 
We will now show,  without assuming that  it is 
supersymmetric  or satisfies the first order equations, 
how to superpose a smooth domain wall background with plane-fronted
gravitational waves moving in the anti-de-Sitter background.

\section{pp-waves on the brane: The bound state}

An exact solution of Einstein's equations
representing a gravitional wave moving at the speed of light in the 
$x^1$ direction is given by retaining the form $\Phi(r)$ and $A(r)$ but
modifying the metric (\ref{metI}) to take the form:
\ben
ds^2= dr^2 + e^{2A(r)} \Bigl ( -dudv + H(u, r, x^i_{\perp}) du^2 + 
dx^i_{\perp} dx^i_{\perp}  \Bigr ) ,\label{metII}
\een
with $u=t-x^1$, $v=t+x^1$, $i=2,\dots ,d-3$ and where the $u$ dependence
of $H$ is arbitrary but it's dependence upon $r$ and $x^i_{\perp}$ is 
governed
by
\ben
H^{\prime \prime} +(d-1) H^\prime A^\prime + 
e^{-2A} \nabla ^2_{\perp}H=0, \label{Heq}
\een
where $\nabla ^2_{\perp}$ is the flat Laplace operator
in the  coordinates $x^i_{\perp}$. This will have half as much
supersymmetry as the domain wall background.  One 
may further generalize this solution by replacing the 
flat metric $dx^i_\perp dx^i_\perp$ by an arbitary 
$(d-3)$-dimensional Ricci flat metric $g_{\perp}$.  If
$g_{\perp}$ admits covariantly constant spinors, then the
background will still admit some supersymmetry.

If $g_{\perp}$ is flat space, solutions of (\ref{Heq}) propagate
in surfaces of constant $r$ at the speed of light in the 
(arbitrarily chosen) $x^1$ direction with an amplitude depending
upon $r$.  Fourier analyzing in the $x_{\perp}$ direction
gives $H \propto e^{ik {\cdot} x_{\perp}}$, where $k$ could in
principle depend upon $u$.  If $k$ is real, solutions would propagate
faster than light in a given $r=$constant surface, and would appear
as tachyons to an observer on the brane.  On the other hand,
solutions for which $k$ is pure imaginary propagate on the brane
like Kaluza-Klein modes.  Thus, if $k^2 = -m^2$, i.e.,
${\nabla ^2}_{\perp}H = m^{2}H$, we are led to the equation
\ben
H^{\prime \prime} +(d-1) H^\prime A^\prime + 
e^{-2A} m^{2}H=0. \label{Heq1}
\een

Consider the zero modes, i.e., solutions with $m^2 = 0$.
We take $H= F(r)H_{ij}(u) x^i_{\perp} x^j_{\perp}$ and find
that $F = C_1 + C_{2}\int^r dse^{-(d-1)A(s)}$ where $C_1$ and
$C_2$ are constants.  The graviton 
perturbation $h= e^{-2A} H$ will diverge exponentially
for large values of $|r|$ unless $C_2 = 0$.  We will return
to this divergence in the next section.  The mode for which
$C_2 = 0$ and $C_1 = 1$ and 
\ben
H= H_{ij}(u) x^i_{\perp} x^j_{\perp} \label{Heq3}
\een

\noindent may be identified as a fully non-linear
version of the zero mode of Randall and Sundrum on a general
domain wall background.  Here, 
$H_{ij}(u)$ is an arbitrary trace free symmetric matrix which
determines the polarization state of the graviton.
The choice (\ref{Heq3}) is made so that the solution has
a d-dimensional isometry group acting on the surfaces 
$r=$constant, $u=$constant.  This invariance is not manifest
in the coordinates $(r,u,v,x_{\perp})$, but is in Rosen coordinates
\cite{gazza}
$(\tilde u, \tilde v, \tilde {x_{\perp}})$,
in which (\ref{metII}), given (\ref{Heq3}), assumes the form
\ben
ds^2 = dr^2 +e^{2A}(-dud\tilde v + A_{ij}(u)d{{\tilde x}_{\perp}}^{i}
d{{\tilde x}_{\perp}}^{j})
\een

\noindent where $u = \tilde u$, $v = \tilde v + 
{1 \over 2}{\dot A }_{ij}(u){{\tilde x}_\perp}^i{{\tilde
x}_\perp}^j$, and ${x_\perp}^i = 
P^i {\thinspace} _j(u){\tilde x}_\perp^j$.  Here,
$A_{ij}(u) = P^m {\thinspace} _i(u)P^m {\thinspace} _j(u)$, $\cdot$
denotes differentiation with respect to $u$ and the matrix
$P^{i} {\thinspace} _{j}(u)$ is a solution of 
${\ddot P}^{i} {\thinspace} _{j} = H_{ik}P^k {\thinspace} _{j}$.
To make contact with (\cite{rs1},\cite{rs2}), we linearize, setting
$P_{ij} = {\delta}_{ij} + \frac{1}{2} {\psi}_{ij}$ so that 
${\ddot \psi}_{ij} = H_{ij}$.  The quantity $\psi$ is essentially
the perturbation considered in (\cite{rs2}).  Rosen coordinates
are in general rather pathological at the non-linear level and 
awkward to use.  In our non-linear analysis we shall, from
now on, only use the coordinates $(r,u,v,x_{\perp})$.

\section{pp-waves in the bulk: Blueshift and Curvature 
Singularities}

Our spacetimes are timelike and lightlike geodesically incomplete as 
$|r| ~{\longrightarrow}~ \infty$.  In the absence of
gravitational waves, i.e., $H=0$, $r = \infty$ corresponds
to a regular Cauchy horizon, and the solution may be extended
through the horizon (see for example \cite{walls}).  
If $H {\neq} 0$ however, the solutions will generically become
singular as $|r| ~\longrightarrow~ \infty$, and will not
admit an extension.  The nature of this singularity 
is most easily studied when the background is taken to be exactly
$AdS_d$.  If we let $z = ae^{r/a}$ then the metric (\ref{metII})
can be recast in so-called `Siklos' coordinates \cite{jiri1}:
\ben
ds^2 = \frac{a^2}{z^2}(dz^2 -dudv + Hdu^2 + 
dx^i_{\perp} dx^i_{\perp}), \label{sikmet}
\een

\noindent where $H$ now satisfies the generalized Siklos equation
\[
z^{(d - 2)}\frac{\partial}{{\partial}z}
[\frac{1}{z^{d - 2}}\frac{{\partial}H}{{\partial}z}] +
{\nabla}^{2}_{\perp} H = 0.
\]

\noindent Because all invariants formed from the Weyl tensor of
(\ref{sikmet}) necessarily vanish, it is not possible to detect
curvature singularities directly by calculating invariants.
However, the necessary condition that one may extend through the singularity
in the metric at $z = {\infty}$ is that the components
of the Riemann tensor in an orthonormal frame which has 
been parallelly propagated along every timelike geodesic
are finite.
This requirement arises because freely falling observers move along timelike
geodesics, and the components of the 
curvature tensor will measure the tidal forces
which these observers experience.  Following the demonstration
in \cite{jiri1}, one may calculate these terms explicitly for
the Siklos metrics.  One finds that certain frame components of
the Riemann tensor generically assume the form
\ben
R_{(a)(b)(a)(b)} = \frac{\Lambda}{d-1} {\pm} z^5(\frac{1}{z}
\frac{\partial H}{\partial z})_{,z} \label{Riem}
\een

\noindent where we have suppressed various constants which are
irrelevant to this discussion.  It follows that any solution
with $z$-dependence cannot be extended, and hence is singular.
One sees that the z-dependent
piece of (\ref{Riem}) is the contribution from the Weyl
tensor.  It would therefore seem that the gravitons will
be heavily `blueshifted' as we move towards large values of $z$.

If ${\nabla}^{2}_{\perp}H = m^{2}H$, the Siklos equation has solutions
of the form
\ben
H =
z^{\frac{d-1}{2}}e^{ik{\cdot}x_{\perp}}[D_{1}J_{\frac{d-1}{2}}(mz) +
D_{2}Y_{\frac{d-1}{2}}(mz)], \label{siklos}
\een

\noindent where $J_{n}(x)$ and $Y_{n}(x)$ are Bessel functions,
and $D_1$, $D_2$ are some constants.
The $z$-dependence of $H$ has the same form as the Kaluza-Klein modes
of (\cite{rs1}, \cite{rs2}).  The behaviour near $z = \infty$ shows that these
are singular on the Cauchy horizon.

In order to get a better feel for the singular nature of these
spacetimes, it is useful to focus on a specific example of
a Siklos-type metric where the $z$-dependence is non-trivial.
The simplest example is the higher dimensional
generalization \cite {pope} of {\it Kaigorodov's}
spacetime \cite{soviet}, for which $H$ is
\[
H(z) = z^{d-1}.
\]

\noindent The Kaigorodov metric is
\begin{eqnarray}
ds^2 = \frac{a^2}{z^2}(-(1-z^{d-1})dt^{2} 
-2z^{d-1}dtdx^{1} \\
+(1+z^{d-1})(d{x^1})^{2}+dz^{2}+d{x_{\perp}}^2). \nonumber
\end{eqnarray}

\noindent This is the $AdS_d$ analogue of the simplest {\it vacuum}
pp-wave, namely, the homogeneous pp-wave in flat space.
It has $d-1$ obvious translational Killing vectors, and is
also invariant under the ${\Bbb R}^{+}$-action:
\[
(z,u,v) ~\longrightarrow~ 
({\lambda}z,{\lambda}^{\frac{3-d}{2}}u, {\lambda}^{\frac{d+1}{2}}v).
\]

\noindent This action, combined with translations in $u$ and $v$,
generates a three-dimensional group of Bianchi Type $VI_{h}$, where
$h = \frac{-1}{(d-1)^{-2}}$.  Therefore, the Kaigorodov 
isometry group contains
a simply transitive subgroup which takes every point with $z$ positive
to any other point with $z$ positive.  A similar $d$-dimensional simply 
transitive group exists in the $AdS_d$ case, for which the ${\Bbb R}^{+}$
action is simply $z ~\longrightarrow~ {\lambda}z$.
In the $AdS_d$ case, we can extend beyond the reach of the group,
in the Kaigorodov case we cannot.

Clearly, freely falling timelike observers (who can cross
the surface $z = \infty$ after a finite period of affine parameter
time \cite{jiri1}) will see infinite tidal forces in this region.
This shows that there are naked curvature singularities at the
points $z = \infty$.  Given our discussion in the previous section,
where we saw that generic $z$-dependent graviton perturbations
will diverge at large $z$, it is clear that we should regard these
singularities as a generic feature of Siklos spacetimes.

\section{Discussion}

We have shown that it is possible to include a non-linear gravitational
wave on a thick domain wall background, in such a way that one may recover the
Randall-Sundrum bound state. 
Given the formal Witten style stability proofs in \cite{cvetic},
which work as long as one has a solution of the first order equations,
one might have thought that this would ensure that the Randall-Sundrum
scenario could be perturbed in this way without problems.  
However somewhat to our surprise, we have found
that generically gravitons propagating in the bulk become singular
on what is a Cauchy horizon in the unperturbed spacetime.
These singularities are somewhat unusual, in that scalar invariants
formed from the curvature tensor do not blow up but rather the
components of the curvature in a parallelly propagated frame along
a timelike geodesic do blow up.  Such singularities are called
``pp curvature singularities'' \cite{jiri1}.

One might worry that these singularities
signal a breakdown in our ability to make unitary predictions.
However, any statements about unitarity should be restricted
to physics {\it on the brane} at $z=$ constant.  Any pathological
effects which may emerge from the singularity will be heavily 
red-shifted by the time they reach the brane.  
Consequently, the extent to which
these singularities signal a pathology of the theory is 
at present unclear.
Interestingly, if one considers massless $z$-independent pp-waves
(these would correspond to the Randall-Sundrum 
zero mode bound state),
one finds that the components of the curvature do {\it not}
blow up, and presumably the spacetime has a non-singular extension.

To conclude we would like to return to the question
of whether a suitable super-potential exists which can be 
derived from a supergravity model.
The results of \cite{cvetic} and \cite{kallosh} show that  for the simplest case
of a single scalar field in models of the type studied in \cite{paul} they do not.
In fact one may show quite generally that for the models in \cite{paul} with
an arbitrary number scalar fields they do not. The same is true for the models
considered in \cite{gppz}. It therefore remains an important open problem
to find a suitable supergravity model or prove that no such model exists.

The authors thank M. Ba\~nados, J. Harvey, S. Hawking, N. Lambert,
R. Myers,
M. Porrati, H. Reall and S. Siklos for 
useful conversations and correspondence.  A.C. was supported
by Pembroke College, Cambridge.

\end{multicols}

\end{document}